\documentclass[aip,rsi,reprint,floatfix]{revtex4-2}  
\usepackage{graphicx}  
\usepackage{dcolumn}   
\usepackage{bm}        
\usepackage{amssymb}   
 \usepackage{float}
\usepackage{amsfonts}
\usepackage{gensymb}
\usepackage{amsmath}
\usepackage{upgreek}
\usepackage{svg}
 \usepackage{textcomp}
 \usepackage{hyperref}

\hypersetup{
    colorlinks=true,
    linkcolor=blue,
    filecolor=cyan,      
    urlcolor=cyan,
}
 
\usepackage{listings}
\usepackage{color}

\definecolor{dkgreen}{rgb}{0,0.6,0}
\definecolor{gray}{rgb}{0.5,0.5,0.5}
\definecolor{mauve}{rgb}{0.58,0,0.82}

\usepackage{blindtext,tikz}
\usetikzlibrary{calc}
\nocite{*}

\begin{document}
\title{Edge Detection and Image Filter algorithms for Spectroscopic Analysis with Deep Learning Applications}
\author{Christopher Sims}
\thanks{Sims58@purdue.edu}
\affiliation{\textit{School of Electrical and Computer Engineering, Purdue University, West Lafayette, Indiana 47907, USA}}
\date{\today}
\begin{abstract}

Edge detection and image filters are commonly used in computer vision. However, they have never been applied to the data analysis of angle resolved photoemission spectroscopy (ARPES) data before in a systematic fashion. In this paper we will use the Sobel, laplacian of a gaussian (LoG), Canny, Prewitt, Roberts, and fuzzy logic methods for edge detection in the ARPES results of HfP$_2$, ZrSiS, and Hf$_2$Te$_2$P$_2$. We find that the Canny filter is the best method for edge detection of noisy data that is typical of ARPES measurements, while the other edge detection techniques are not able to correctly detect ARPES bands. 
\end{abstract}
\maketitle

\section{Introduction}
As spectroscopic technology has advanced over the years, angle resolved photoemission spectroscopy (ARPES) has become a more prominent tool in classifying the properties of topological quantum materials\cite{Damascelli2004,Comin2015,Lu2012}. Through ARPES, the electronic structure can be viewed in momentum space revealing their electronic properties. In addition ARPES plays a critical role in the classification of topological quantum materials such as Dirac \cite{Hasan2010.Xia2009a}, Weyl \cite{Xu2015,Lv2015,Huang2015}, and nodal line semimetals \cite{Burkov2011,Fang2015,Fang2016}. With the advancements in ARPES technologies such as pump-probe, and spin resolved ARPES, there exists even bigger data sets that need to be analyzed systematically. There is a symbiotic connection between ARPES and density fucntional theory (DFT) band structure calculations \cite{Wu2018,Giannozzi2009,Gonze2009,Mostofi2008} in which theory is able to be confirmed experimentally improving both techniques and deepening our understanding of topological quantum materials. To do this, accurate analysis of data from ARPES and other techniques is necessary but not always possible due to noisy data as a result of poor sample quality or non-optimal experimental parameters. Through the use of edge detection and image filtering, key features can be highlighted and noise can be suppressed further revealing fetures that are difficult to see by eye. In cases where there is noise, data can be made significantly easier to trace by eye and made coherent through edge detection techniques. These methods work by taking advantage of the concept the curl which is large near ares of large change. By approximating derivatives at individual pixels of the image and finding maximal points of change (i.e. the edge), a new image consisting of that approximation at each point can be created in which the background is suppresed and only the edges are shown. To do this, simple operators such as the Sobel operator choose a kernel, or a matrix consisting of a chosen center point and its surrounding pixels, and discretely differentiate at each point, by doing this separately in the horizontal and vertical directions you can approximate the gradient magnitude and direction. The approximate gradient can give a proposed direction and magnitude of and edge, and by applying thresholds, a new image is formed. This is the general premise of edge detection operators and can be done with higher order derivatives as well, with each subsequent order sharpening the definition of edges but consequently amplifying noise, in especially noisy images and data sets, edges can’t be discerned without somehow suppressing the noise before applying an edge detection operator. To combat noise in image filtering a blurring operator is commonly applied that works to “smooth” the image and nullify high frequency variation that occurs in noisy areas of an image. By filtering an image before applying the derivative operators, to some degree, the noise can be suppressed and allow for the previously discussed operation of edge detection to be applied with greater success.

In this paper we cover the effectiveness of the Sobel\cite{Gao2010,Ma2010}, Prewitt\cite{Senthilkumaran2009}, Roberts \cite{G.T.Shrivakshan}, Canny\cite{Canny1986}, Laplacian of (a) Gaussian (LoG)\cite{Huertas1986}, and the fuzzy logic filters. We apply these methods to both pure edge detection algorithms and image filter algorithms using the \emph{SpectroLab} suite of programs and \emph{MATLAB}’s built in functionality. With the analysis tools offered through the \emph{SpectroLab} suite, utilizing deep learning, the researchers’ task of data analysis can be further simplified. By supplying a wide variety of classified data to a machine learning program, through a process known as unsupervised image exploration, a program can become proficient at image segmentation and object tracking \cite{Peng2020,Xiao2018,Molini2019,Hui2018,Pandey2018,Mou2018,Kim2018,Zangeneh2017,Zhang2018,Wang2016}. In conjunction with theory, a program could completely classify the band structure and other features of a sample available from the given data set, relatively quickly by extrapolating certain individual points that correlate well with known data sets. However, in order to do this, a large data set of well formatted ARPES and DFT must be created in order to fully implement these machine learning algorithms.

\section{Methods}
Using the \emph{MATLAB} based program \emph{SpectroLab}, we have integrated a code to test all of the edge detection and image filters using \emph{MATLAB}'s builtin tools.
\section{Image Filters}
\subsection{Sobel filter}
The common edge detection algorithms are the Sobel, Canny, Prewitt, Roberts, and fuzzy logic methods. Here we use a modified version of the Sobel filter in order to conduct second derivative calculations. The Sobel filter is a 3$\times$3 matrix that calculates the second derivative locally in an image file (where A is the image). There are two versions, the horizontal and vertical which can be used, in the distributed program the horizontal $F_x$ is convoluted with the image to form the second derivative of the image.
\begin{equation}
G_x  = 
  \begin{bmatrix}
    -1 & 0 & +1 \\
    -2 & 0 & +2\\
      -1 & 0 & -1
   \end{bmatrix} * A , 
   G_y =
     \begin{bmatrix}
    -1 & -2 & -1 \\
    0 & 0 & 0\\
      +1 & +2 & +1
   \end{bmatrix} * A
\end{equation}
Once the horizontal and vertical derivative approximations are formed, the gradient can be approximated by finding the manitude and direction of the approximate gradient at each point. Once this is done, direction of edges can be discerned and classified based on thresholds applied to the calculated magnitudes.

\subsection{Prewitt}
The Prewitt filter uses a 3$\times$3  kernel which is convoluted with the image to calculate the derivative for the horizontal and vertical direction. Again, we define A as the source image and $G_x$ and $G_y$ as the gradient filters.
\begin{equation}
G_x  = 
  \begin{bmatrix}
    -1 & 0 & +1 \\
    -1 & 0 & +1\\
      -1 & 0 & +1
   \end{bmatrix} * A , 
   G_y =
     \begin{bmatrix}
     +1 & +1 & +1 \\
      0 & 0 & 0\\
      +1 & +1 & +1
   \end{bmatrix} * A
\end{equation}
The x and y gradients can be combined to calculate the total gradient at each point with
\begin{equation}
G = \sqrt{G_x^2 + G_y^2}
\end{equation}
	Unlike the Sobel filter, the Prewitt filter is "more isotropic" and depending on which kernel is used (Horizontal or Vertical), will emphasize diagonal changes as much as the designated ones for that specific kernel. This can be useful but also decreases the usefulness of using specific kernels for the horizontal and vertical approximate derivatives and can result in amplification of noise and deviations in the gradient calculations.

\subsection{Laplacian of a Gaussian (LoG)}
The Laplacian of a Gaussian (LoG) filter uses an alternate filter which is dependent on the Gaussian ($\sigma$) that one wishes to use, here we provide two examples of the (LoG) filter
\begin{equation}
G_1 = 
  \begin{bmatrix}
    0 & -1 & 0 \\
    -1 & 4 & -1\\
      0 & -1 & 0
   \end{bmatrix} * A , 
   G_2 =
     \begin{bmatrix}
     -1 & -1 & -1 \\
      -1 & 8 & -1\\
      -1 & -1 & -1
   \end{bmatrix} * A
\end{equation}

	Note that for the applications of analysis of ARPES data, using a LoG filter can be especially useful in suppressing background noise but because of the isotropic nature of the Gaussian filter, can have negative effects on the distinct band structures and can even make them indiscernible as shown in figure. If, instead, an anisotropic blurring filter was applied, this would not be the case although the process would be more computationally expensive and complex in general.

\subsection{Roberts}
The Roberts filter is dissimilar to typical gradient algorithms because it takes diagonal gradients of the image using the following kernel filters.
\begin{equation}
G_1 = 
  \begin{bmatrix}
    +1 & 0 \\
    0 & -1
   \end{bmatrix} * A , 
   G_2 =
     \begin{bmatrix}
      0 & +1 \\
      -1 & 0 
   \end{bmatrix} * A
\end{equation}

	This algorithm is notable because of it's relative simplicity and efficiency. Rather than detecting lines as edges as gradient operators such as the Sobel and other variants of it attempt to do, the Roberts operator is a point detector and works to classify edges by choosing individual pixels which differ significantly from those around it, as a result, the Roberts filter is highly sensitive to noise and a salt and pepper filter applied to an image or graininess in general can greatly reduce the abilities of the Roberts filter. At the cost of information on edge direction and accuracy, the Roberts filter is the most efficient image filter, especially for use on binary images.

\subsection{Canny} 
The canny filter works be first applying a Gaussian filter to smooth the image, the finding the gradients and suppressing noise, apply a threshold, and finally track the edge via hysteresis. By calculating the angle of the gradient with the equation $\Theta = tan^{-1}(G_y,G_x)$ on can suppress the edges using a set of rules defined by the Canny filter
\begin{itemize}
\item $\Theta = 0\degree$ will be an edge if the magnitude is greater the the left and right pixels.
\item $\Theta = 90\degree$ will be an edge if the magnitude is greater than the up and down pixels
\item $\Theta = 135\degree$ will be an edge if the magnitude is greater the the top-right and bottom-left pixels ($\searrow$ or$ \nwarrow$)
\item $\Theta = 45\degree$ will be an edge if the magnitude is greater that the top-left or bottom-right pixels ($\nearrow$ or$ \swarrow$)
\end{itemize}
Using blob analysis (or another hysteresis algorithm) edges are tracked to see if they are true edges. These edges are either deleted or connected depending on the blob analysis.

\section{Edge Detection on ARPES data}
	As previously discussed, operators such as the Sobel and Prewitt operators will approximate edges through gradient approximation, as a result there is both a horizontal and vertical SDI created by the operator. Fig. 1 and 3 A-E and F-J (first and second rows) are the horizontal and vertical approximations respectively and the differences in each are well illustrated. As expected, in the horizontal derivative approximations, there is a distinct bias towards horizontal changes and edges, evidenced by clearer and harder edges for these horizontal changes, and similarly for the vertical derivative approximations. 

	The exception to this rule is the instances where the Laplacian operator is applied, these images look identical in both rows [Fig. 1 A-E, F-J] because due to the nature of the Laplacian operator there is no distinction between the horizontal and vertical discrete derivatives as it approximates the edges by finding local maximums via the second derivative method. A second derivative, or Laplacian, of the image is found and zero crossings are identified in order to classify edges. By using the isotropic Laplacian operator, rather than the anisotropic Sobel, Prewitt, and Roberts operators, any bias towards certain edge orientations is eliminated and similarly, the orientation of the data will pose no changes to the analysis. In general, by taking the second derivative approach of the Laplacian operator, stronger edges are emphasized but high frequency noise is amplified considerably. This means that some type of blurring, typically through the use of a Gaussian convolution, is necessary. In especially noisy data sets, applying such a method can be effective in reducing background noise and detecting band structure but by applying an isotropic filter such as the Gaussian, the key features such as band structure will also be obscured [Fig. 5]. Fig. 2 and 4 show the use of a low $\sigma$ valued Gaussian filter with small filter size in conjunction with the Laplacian, compared with only applying the Laplacian the results are the same. For particularly clean data such as this, applying a substantial amount of blurring is detrimental to the process of image analysis [Fig. 5]. 
	
	Not only does the Laplacian differ in the sense that it uses second derivative techniques to classify edges, but because of the use of a single isotropic kernel, it is a relatively more efficient mask to apply than other gradient based operators while not sacrificing robustness as the Roberts operator does. In general though, there are cases in which the basic operators such as the Sobel, Prewitt, and Canny algorithms are preferable to the Laplacian algorithm, as seen in the better definition in images produced by the Sobel and Prewitt operators [Fig. 1 and 3]. 
	
	Fig. 2 and 4, (A-F) and (G-L), are the \emph{MATLAB} and \emph{SpectroLab} implementations of various edge detection operators, respectively. To begin with, each of these implementations are now the full, gradient approximations, of course excluding the Laplacian operators [Fig. 2 \& 4, E,J,K]  which utilize the same isotropic operator as in [Fig. 1 and 3]. Rather than approximating edges through horizontal or vertical discrete derivatives, these images show an implementation which takes both of the horizontal and vertical approximations and uses them to create a gradient approximation through the formulas described in the methods subsection. In this case, the Sobel and Prewitt operators [Fig. 2 \& 4, H \&I] are essentially identical due to the small differences between the two operators, ie the increased emphasis on horizontal or vertical changes in the horizontal and vertical approximations of the Sobel operator's implementation. It is for this reason that we can say that it may be acceptable to use either one, interchangeably, for the purpose of ARPES data analysis, due to the Prewitt operator being a simpler mask to apply it could be preferred over the Sobel operator. Similarly to the interchangeability of the Sobel and Prewitt operator for ARPES data analysis, the Laplacian of a Gaussian(LoG) and Laplacian operators [Fig. 2 \& 4, J \& K] yield identical results, as mentioned before, this is due to the use of a low $\sigma$($\sigma$ = 1) value for the gaussian smoothing in the implementation of the LoG operator. The effect of using a low $\sigma$ value and filter size is shown when comparing [Fig. 2 \& 4, G \& L], the Gaussian filtered image looks identical to the original. This would not be the case if a larger $\sigma$ value or filter size is used as a more extreme smoothing effect would be seen which is desirable in cases of higher noise as seen in [Fig. 5]. Seeing as we have relatively clean data, the use of a higher $\sigma$ value or filter size would cause for less definition in edges that are already not optimal compared to the Sobel and Prewitt results [Fig. 2, 4 \& 5]. Although the LoG filter is highly useful in analysis of normal images, we can see that applying a substantial $\sigma$ value can be detrimental to our goals of classifying band structures, if the data was significantly blurred, we would not be able to see the double bands depicted in the filters that do not utilize blurring in [Fig. 4] and thus, a simple Laplacian operator should be favored over a LoG excepting in cases of quite high frequency noise.

	In general, the built-in \emph{MATLAB} implementations are not very useful in the context of analyzing ARPES data and both amplify noise, considerably in the case of the LoG and Canny operators, while not yielding any figure with discernible edges that could be useful in interpreting ARPES data. Juxtaposed with the \emph{SpectroLab} implementations which work to yield images that much closer resemble the original while highlighting the important band structures. Although the formulation of custom edge detection operators for the purpose of ARPES data analysis would be ideal, these standard image analysis operators work well on ARPES data and seem to be suitable replacements for current second derivative image analysis techniques.
	
\section{Outlook}
With the ability to now use edge detection and image filters on ARPES data, we open up to the possibility of using machine learning and neural networks in order to reverse engineer topological materials. With the indexing of a large number of already discovered topological materials with machine learning we may soon be able to engineer new materials using deep learning algorithms.

Using these image analysis algorithms, we are currently developing a machine learning algorithm to index experimentally discovered topological materials with the aim to predict the band structure of an already discovered topological material to prove the effectiveness of the algorithm. After training the algorithm, we aim to predict new topological quantum materials.
\vspace{-12pt}
\section{Conclusion}
In conclusion, we have covered most of the common edge detection algorithms and explored their use in the analysis of ARPES data. We find that the canny filter is good in finding edges however a better edge detection algorithm needs to be designed for good data analysis. These filters are good for image recognition algorithms, however, they are also excellent to replace current second derivative analysis algorithms of ARPES data. With the ability to use image filters and edge detection we open up the ability for advanced techniques such as deep learning and generative adversarial neural networks (GANs) to be applied to topological quantum materials.
\vspace{-12pt}
\section*{Acknowledgments}
C.S. acknowledges the generous support of the Purdue ECE ASIRE fellowship and the National GEM consortium fellowship\\

Correspondence should be addressed to C.S. (Email: Sims58@purdue.edu).
%

\newpage

\begin{figure*}[!ht]
\begin{center}
\includegraphics[width = 1\textwidth]{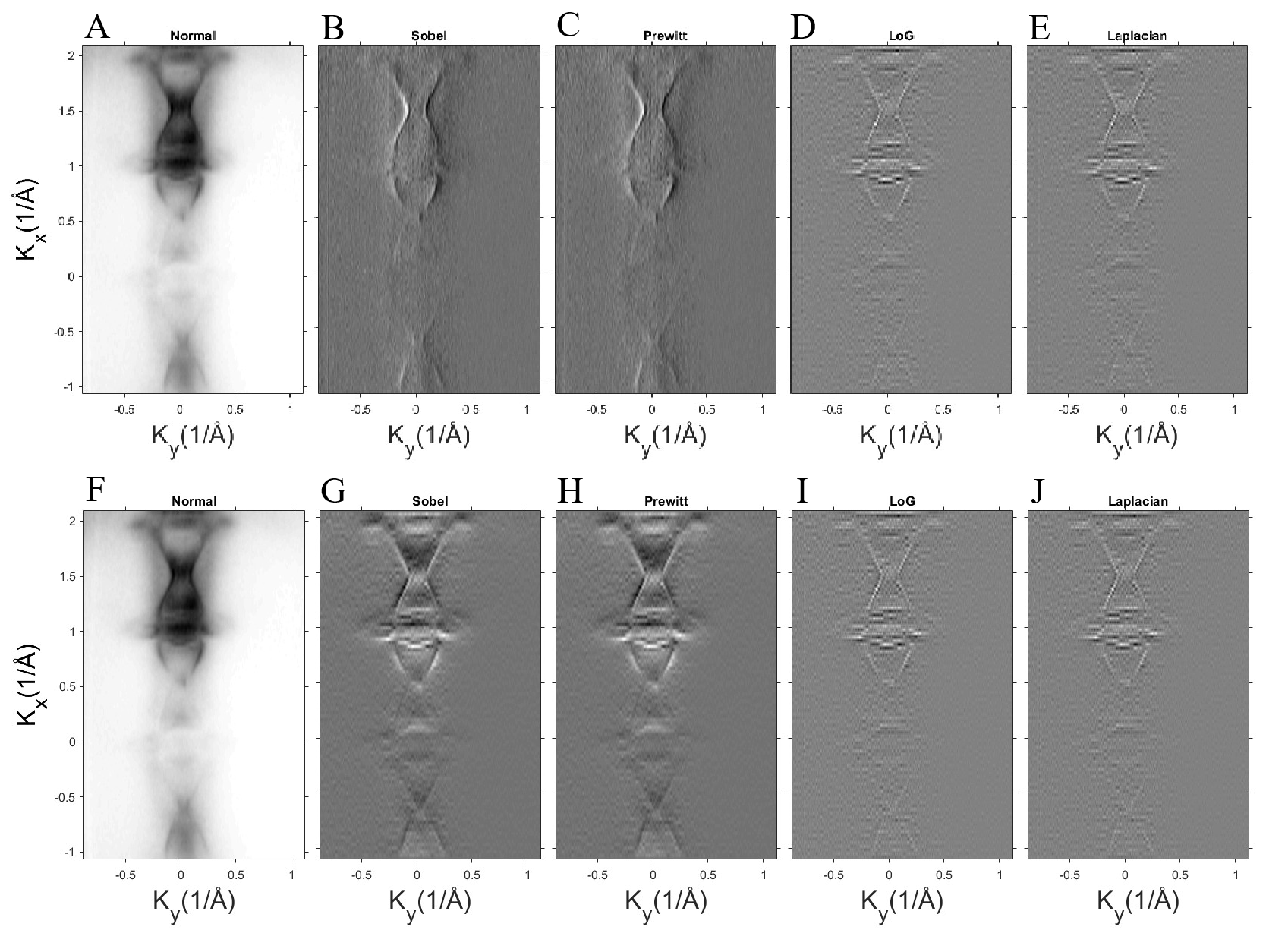}
\end{center}
\caption{\textbf{Implementations of individual horizontal and vertical 1$^{st}$ derivative, and isotropic Laplacian operators.} A,F) The original image of HfP$_{2}$ ARPES data, taken near the Fermi level. B-E) \emph{SpectroLab} implementations of various edge detection operators in the horizontal, k$_{y}$, direction.  G-J) \emph{SpectroLab} implementations of various edge detection operators in the vertical, k$_{x}$, direction.}
\end{figure*}

\begin{figure*}
\begin{center}
\includegraphics[width = 1\textwidth]{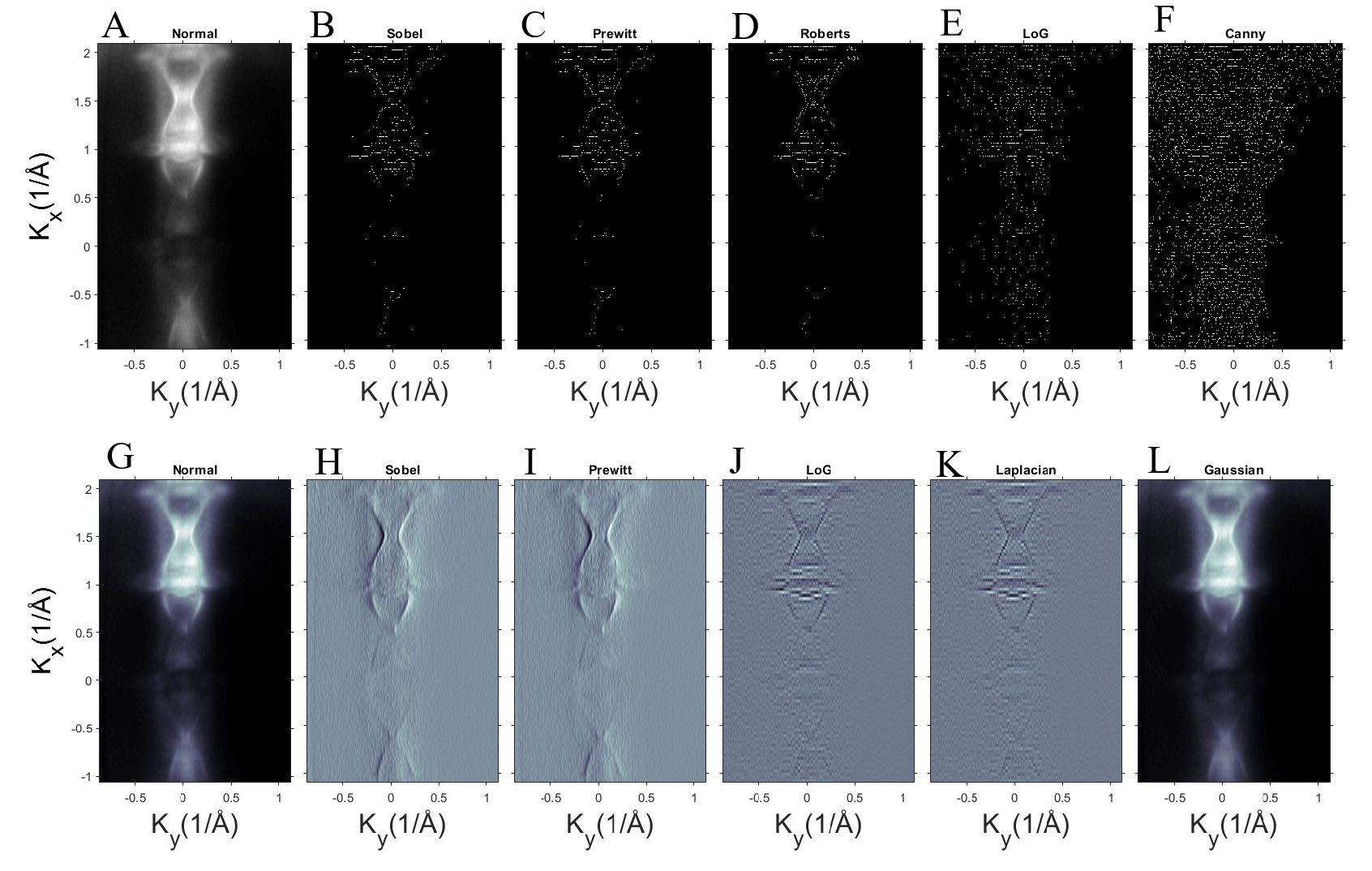}
\end{center}
\caption{\textbf{ARPES data of HfP$_{2}$ at the (001) surface near the Fermi level with  implementations of different image filters in the top and bottom rows, respectively.} (A,G) Original image, \emph{MATLAB} implementation using black/white coloring only. (B-F) \emph{MATLAB} implementations of Sobel, Prewitt, Roberts, LoG, and Canny operators respectively. (H-K) implementations of Sobel, Prewitt, LoG, and Laplacian operators. (F,L) Normal image with a Gaussian filter of $\sigma=1$ and small filter size applied.}
\end{figure*}

\begin{figure*}[!ht]
\begin{center}
\includegraphics[width = 1\textwidth]{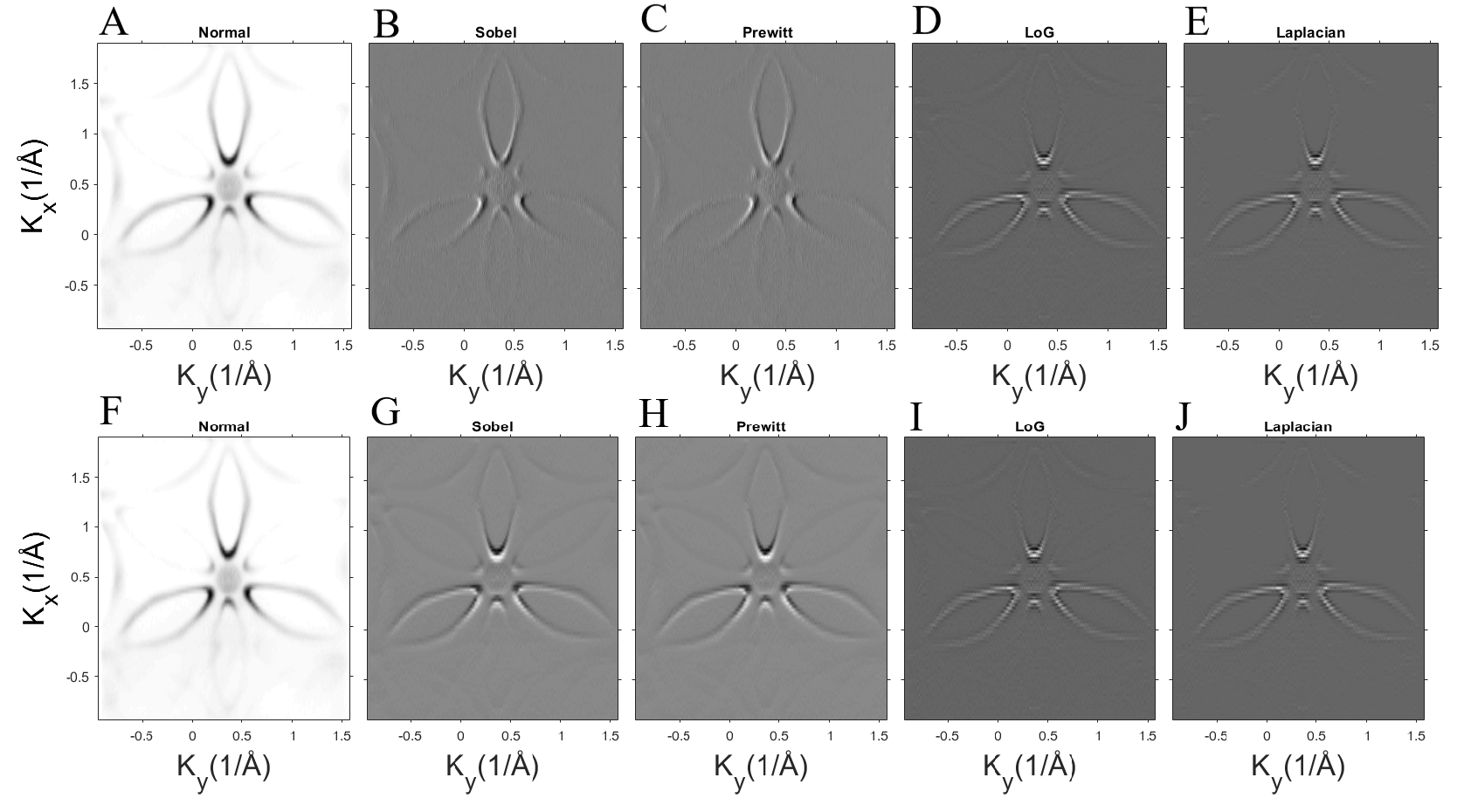}
\end{center}
\caption{\textbf{Implementations of individual horizontal and vertical 1$^{\mathrm{st}}$ derivative, and isotropic Laplacian operators.} (A,F) The original image of HfP$_{2}$ ARPES data, taken near the Fermi level. (B-E) implementations of various edge detection operators in the horizontal, $k_{y}$, direction.  (G-J) implementations of various edge detection operators in the vertical, $k_{x}$, direction.}
\end{figure*}

\begin{figure*}[!ht]
\begin{center}
\includegraphics[width = 1\textwidth]{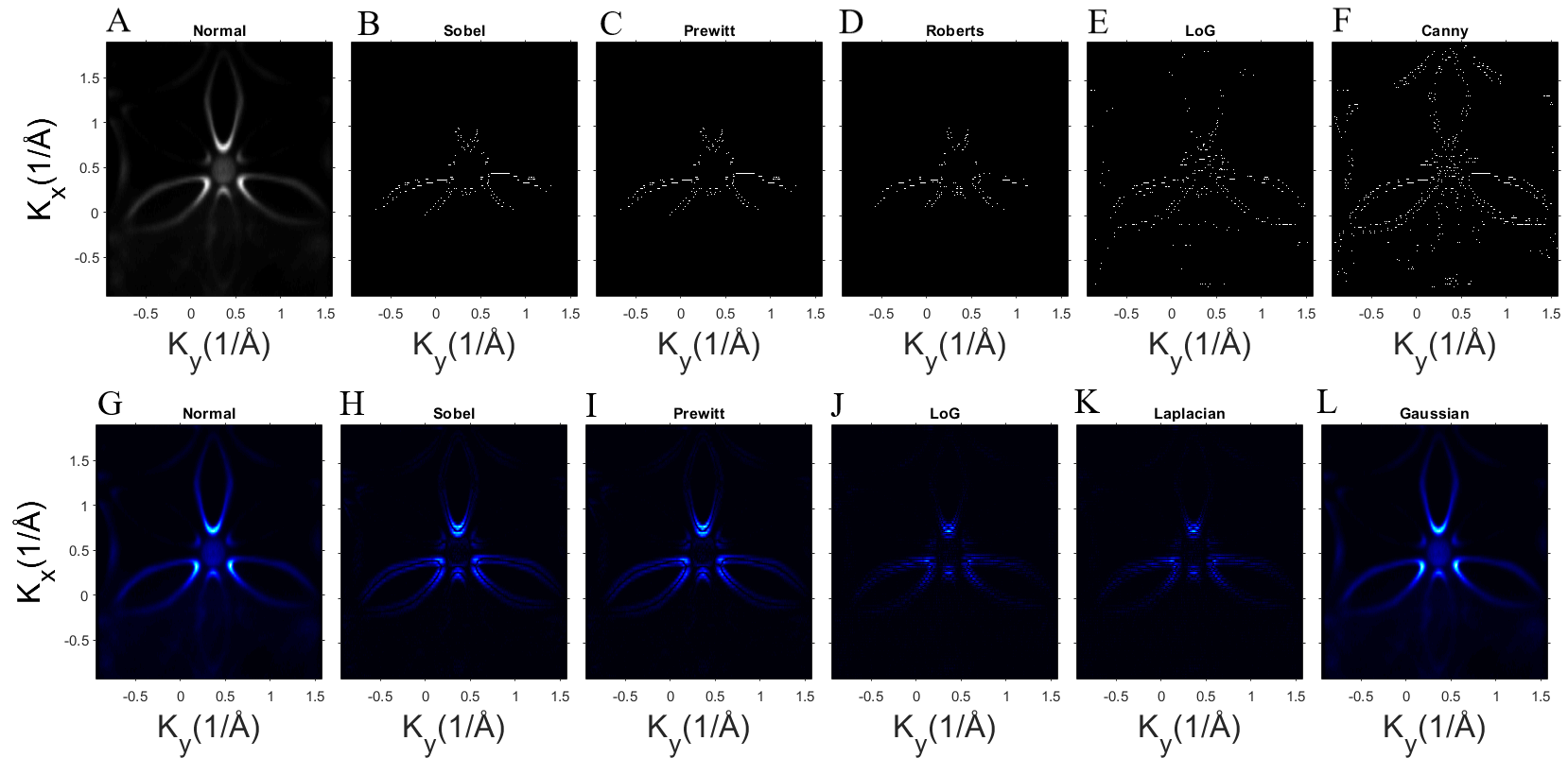}
\end{center}
\caption{\textbf{ARPES data of HfTe$_{2}$P$_2$ at the (001) surface near the Fermi level with implementations of different image filters} (A,G) Original image, \emph{MATLAB} implementation using black/white coloring only. (B-F) implementations of the full Sobel, Prewitt, Roberts, LoG, and Canny operators respectively. (H-K) implementations of the half Sobel, Prewitt, LoG, and Laplacian operators. (L) Normal image with a Gaussian filter of $\sigma=1$ and small filter size applied.}
\end{figure*}

\begin{figure*}[!ht]
\begin{center}
\includegraphics[width = 1\textwidth]{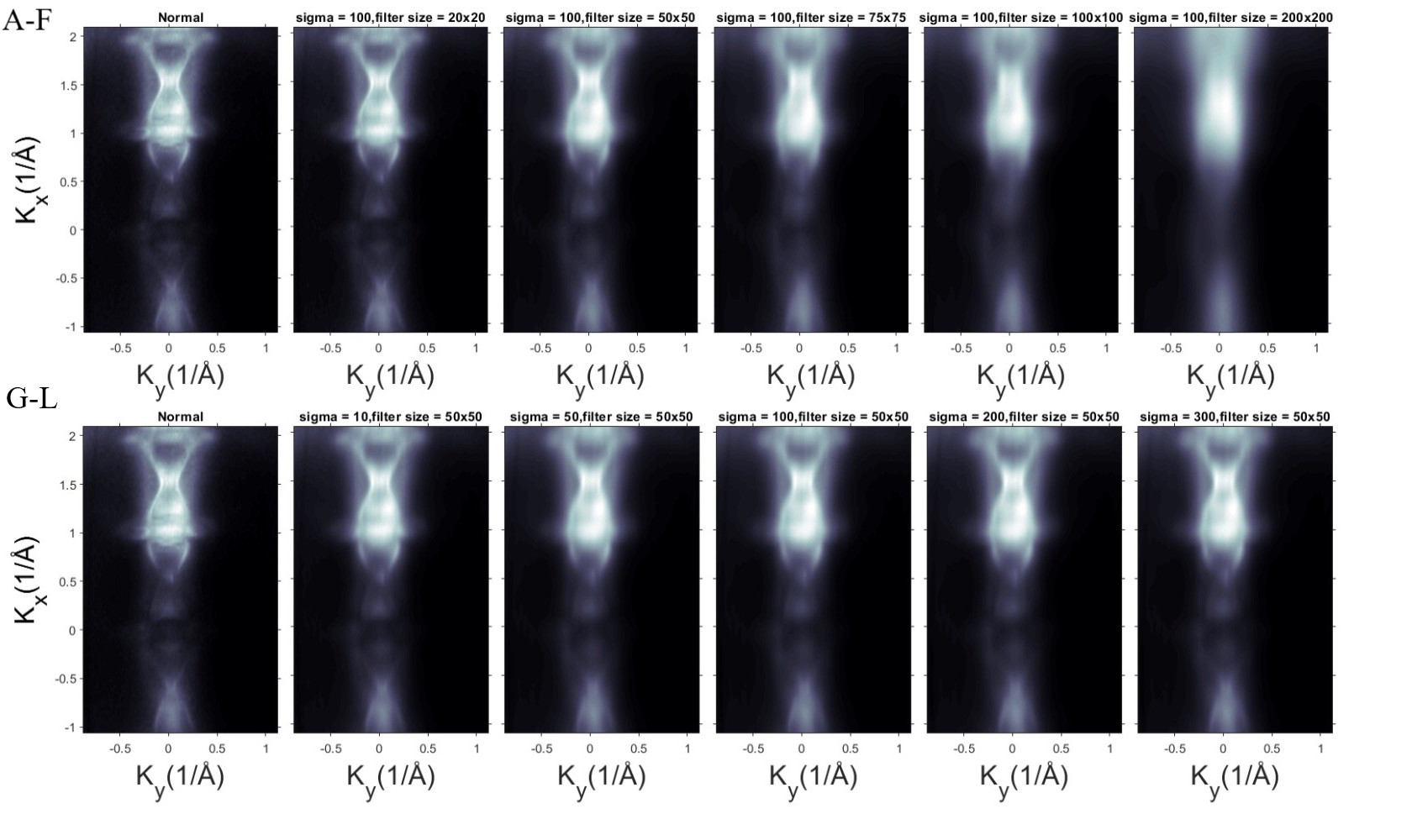}
\end{center}
\caption{\textbf{The effects of various sigma values and filter sizes for the specific Gaussian applied to ARPES data of HfP$_{2}$ at the (001) surface near the Fermi level} (A,G) Original figure with no Gaussian blurring applied. (B-F)Gaussian smoothing is applied with increasing filter size and constant $\sigma$ value. (H-L) Gaussian smoothing applied with increasing $\sigma$ value and constant filter size.}
\end{figure*}

\end{document}